\documentclass[runningheads]{cl2emult}

\usepackage{makeidx}  
\usepackage{graphicx} 
\usepackage{subeqnar} 
\usepackage{multicol} 
\usepackage{cropmark} 
\usepackage{eso}      
\makeindex            

\begin{document}
\title*{Mining the local Universe: the QSO space density
}
\toctitle{Mining the local Universe: the QSO space density}
%
%
\titlerunning{The QSO space density }
%
\author{Stefano Cristiani\inst{1,2}
\and Andrea Grazian\inst{2,1}
\and Alessandro Omizzolo\inst{3}
\and Christopher Corbally\inst{3}
}
\authorrunning{Stefano Cristiani et al.}
%
%
\institute{ST European Coordinating Facility, European Southern
Observatory, Karl-Schwarzschild-Strasse 2, D-85748 Garching bei
M\"unchen, Germany 
\and Dipartimento di Astronomia,
Vicolo dell'Osservatorio 5, I-35122 Padova, Italy
\and Vatican Observatory Research Group, University of Arizona,
Tucson AZ 85721, US
}

\maketitle              

\begin{abstract}
We present progress results of a new survey for bright QSOs
($V<14.5$, $R<15.4$, $B_J<15.2$) 
covering the whole sky at high galactic latitudes, $|b|>30$.
The surface density of QSOs brighter than $B_J=14.8$ 
turns out to be $2.9 \pm 0.8 \cdot 10^{-3} deg^{-2}$.
The optical Luminosity Function
at $0.04 < z \le 0.3$ shows
significant departures from the standard pure luminosity evolution, 
providing new insights in the modelling of the QSO phenomenon.
\end{abstract}

\section{Introduction}
Why a survey for local QSOs?\\
Because in the epoch of 2dF and SDSS, of the thousands of QSOs at $z\sim 
2$, it is important to have precise information about the spatial
density and the clustering of nearby AGN to provide
zero-point and leverage for the study of the QSO evolution.
As in other fields, it is paradoxical that 
we know much better the properties of the high-z QSO population than
the local one.
The technical problem is however not trivial since, as 
the PG survey (Schmidt \& Green 1983) made clear, it is necessary to
cover the whole sky to obtain sufficient statistics 
where the present data are unsatisfactory:
at $z<0.3$ and magnitudes around $13-15$.
Such a task has become relatively simpler in recent times, thanks to
the availability of large high-quality databases.

\section{The Photometric Database}
What do we need ?\\
First of all, since our goal is the study of the optical luminosity
function (OLF) of QSOs, we need a homogeneous database of optical
fluxes covering the whole sky.
Of various existing possibilities: 
APM\footnote{\tt http://www.ast.cam.ac.uk/$\sim$apmcat/}, 
ROE/NRL\footnote{\tt http://xweb.nrl.navy.mil/www{\_}rsearch/RS{\_}form.html},
USNO1\footnote{\tt http://archive.eso.org/skycat/usno.html},
GSC1\footnote{\tt http://www-gsss.stsci.edu/gsc/gsc.html},
DSS\footnote{\tt http://www-gsss.stsci.edu/dss/dss.html}, 
none turned out to be entirely satisfactory. 
After a careful analysis of their sensitivity,
completeness, accuracy, we have chosen a combination of the catalogs GSC,
USNO and DSS:
\begin{enumerate} 
\item {in the Northern hemisphere objects from the GSC catalog
with $11.0 < V_{GSC} \le 14.5$.
The relation between the $V_{GSC}$ band and
the corresponding Johnson $V$ turned out to be:
$V_{GSC}=V-0.21 $,
with $\sigma_{V}=0.27$ $mag$. 
}
\item{in the Northern hemisphere objects from the USNO catalog with
$13.5 < R_{USNO} \le 15.4$.
The relation between the $R_{USNO}$ band and
the corresponding Johnson-Kron-Cousins $R$ turned out to be
$ R_{USNO}=R_{JKC}+0.096 $,
with $\sigma_{R_{USNO}}=0.27$ $mag$.
}
\item {in the Southern hemisphere we have derived
$B_J$ magnitudes from the Digitized Sky Survey (DSS). Small scans 
(``postage stamps'' of $2' \times 2'$) of each object
of interest and of 20-50 surrounding objects with known GSC $B_J$
magnitudes were extracted from the DSS. The magnitude of the object of 
interest was then calibrated against the GSC objects.
In this way a $\sigma_{B_J}$ of $0.10$ $mag$ was obtained in the interval
$12.0 < B_J <15.5$.
}
\end{enumerate} 
\begin{figure}
\centering
\includegraphics[width=.9\textwidth]{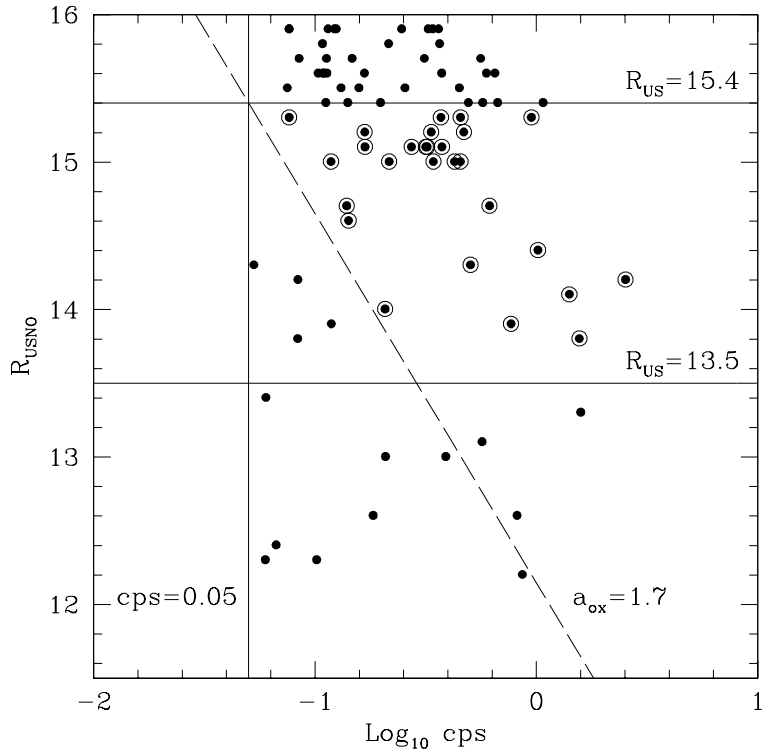} 
\caption[]{The incompleteness in the present selection of QSO
candidates is due to objects not found in the RASS catalogue (with a
flux $\le 0.05$ cps, on the left of the vertical continuous line) and
to those with $\alpha_{ox}\ge 1.7$ (on the left of the dashed line)
\label{fig1}}
\end{figure}
\section{The QSO Selection}
The second step is to select among the millions of objects in the
magnitude range of interest the few hundreds of QSOs. We have used
the RASS Bright Source Catalogue (RASS-BSC, Voges et al. 1999) to
compute the X-optical color,  $\alpha_{ox}$ (La Franca et al., 1995),
which provides a key signature of the AGN phenomenon.
If we take from the V\'eron \& V\'eron catalog (1998) all the non
X-ray selected bright QSOs with $0.04<z<0.3$, assign to them the optical 
magnitudes of our databases and plot them vs. the corresponding RASS 
fluxes, we obtain the diagram of Fig.~1. We can see that, if we select
objects with $\alpha_{ox} < 1.7$ (the circles in Fig.~1), 
the incompleteness does not depend on 
the optical flux: we lose a fraction of about $20\%$ of the QSOs, 
which can be accounted for, 
and we obtain a reasonably short list of candidates:
520 in the North over 8000 sq.deg. and 301 in the South over 5600 sq.deg.

Conveniently, more than $40\%$ of the objects have already an
identification in SIMBAD\footnote{\tt
http://simbad.u-strasbg.fr/Simbad}
or NED\footnote{The NASA/IPAC Extragalactic
Database (NED) is operated by the Jet Propulsion Laboratory,
California Institute of Technology, under contract with NASA}. 
For the remaining ones we are 
taking spectra with the telescopes at the Asiago, La Silla and Kitt Peak
observatories. The success rate is around $50\%$. We have observed $80\%$ 
of the candidates in the North, $50\%$ in the South, collecting a total
of 290 QSOs. At completion we expect to produce a sample of about 450 QSOs.

\section{Results of the Asiago-ESO/RASS QSO Survey}
Preliminary results have been reported in Grazian et al. (2000).
The optical counts have been computed in the $B_J$ band, correcting
our V and R magnitudes according to the average colors of the sample,
and are shown in Fig.~2.
They confirm the well known (Goldschmidt et al., 1992) incompleteness
of the PG survey (about a factor 3) and agree with the extrapolation
of other recent surveys carried out at somewhat fainter magnitudes.
\begin{figure}
\centering
\includegraphics[width=1\textwidth]{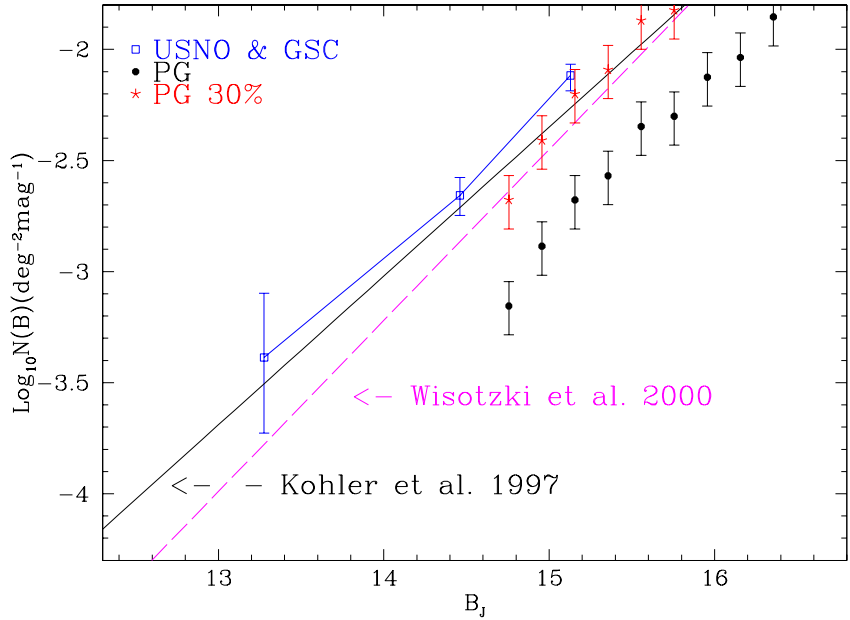} 
\caption[]{The LogN-LogS relation of QSOs. Open squares refer to the
present sample and are QSOs with $z>0.04$. Filled circles show the 
data of PG survey, and stars the same data corrected for a $70\%$
incompleteness. The continuous and dashed straight lines are the
relations found by K\"ohler et al. (1997) and by 
Wisotzki et al. (2000), respectively
\label{fig2}}
\end{figure}
The derived OLF confirms and strengthens the claim of La Franca \&
Cristiani (1997, LC97), i.e. a significant departure of the evolution at
low-z from the pattern generally described with a Pure Luminosity
Evolution (PLE). 
\begin{figure}
\centering
\includegraphics[width=1\textwidth]{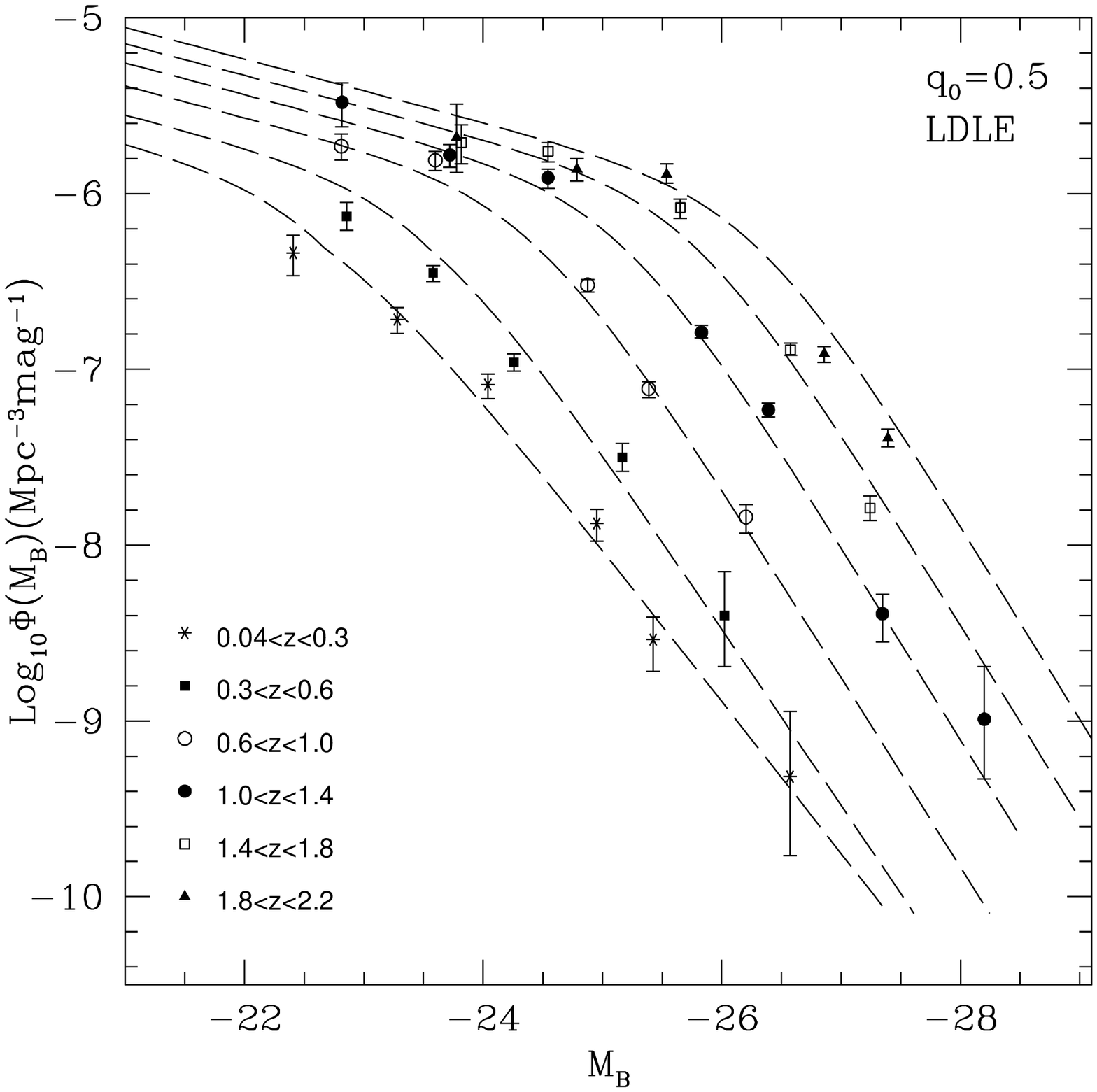}
\caption[]{The luminosity function of QSOs compared with a
parameterization of Luminosity Dependent Luminosity Evolution. 
The points in the range $0.04 < z \le 0.3$ are the result of the
present survey, 
the data in the other redshift ranges are derived from LC97
\label{fig3}}
\end{figure}
It is debatable whether the best parameterization 
of the data is provided by a
Luminosity Dependent Luminosity Evolution (e.g. LC97) or by a
slow-down of the evolutionary rate and/or a mild density evolution at
low-z (e.g. Cavaliere \& Vittorini, 2000).
To obtain a more physical insight it is however advisable to put these
results in the 
context of the models of formation and evolution of 
galactic structures, for evidence is mounting that the QSO
activity, the growth of super-massive black holes (BH) and the formation of 
spheroids are closely linked phenomena.

On the one hand, the increased (with respect to previous measurements)
QSO space density at low-z corresponds better to the predictions of
semi-analytical models (Kauffmann \& Haehnelt 2000). In this scenario
the recurrent activity of short-lived QSOs is driven by the merging
rate of CDM halos, the availability of cold gas and the timescale
of the accretion onto the central BH.  On the other hand, a weak
density evolution and a smoother overall shape of the OLF at low-z are
natural predictions of the model of Cavaliere \& Vittorini
(2000). According to them, the accretion onto an active BH is
controlled by the surrounding structures: efficient fueling is
triggered by the encounters of a gas-rich galaxy with companions in a
group; these destabilize the gas and induce accretion. Strong
luminosity evolution is produced as star formation and
these encounters deplete the gas
supply in the host; an additional, milder density evolution derives,
since the interactions become progressively rarer as the groups grow
richer but less dense.

A refined measurement of the QSO clustering and its evolution, now possible
over nine tenths of the history of the Universe, will be instrumental
in disentangling the various hypotheses.

\clearpage
\addcontentsline{toc}{section}{Index}
\flushbottom
\printindex

\end{document}